\documentclass[conference]{IEEEtran}
\IEEEoverridecommandlockouts

\usepackage{cite}
\usepackage{amsmath,amssymb,amsfonts}
\usepackage{algorithmic}
\usepackage{graphicx}
\usepackage{textcomp}
\usepackage{xcolor}
\usepackage{tikz}
\usepackage{booktabs} 
\usepackage{fontawesome5}
\usetikzlibrary{arrows.meta, positioning, shapes, fit, backgrounds, calc}
\def\BibTeX{{\rm B\kern-.05em{\sc i\kern-.025em b}\kern-.08em
    T\kern-.1667em\lower.7ex\hbox{E}\kern-.125emX}}
\begin{document}

\title{ChipChat: Low-Latency Cascaded \\ Conversational Agent in MLX}

\author{\IEEEauthorblockN{Tatiana Likhomanenko$^*$, Richard He Bai$^*$, Zijin Gu$^*$, Zakaria Aldeneh$^*$, Shiladitya Dutta$^{*}$, \\ Luke Carlson$^{*}$, Han Tran$^*$, Yizhe Zhang, Ruixiang Zhang, Huangjie Zheng, Navdeep Jaitly$^*$\thanks{$^*$Core contributors.} \thanks{© 2025 IEEE.  Personal use of this material is permitted.  Permission from IEEE must be obtained for all other uses, in any current or future media, including reprinting/republishing this material for advertising or promotional purposes, creating new collective works, for resale or redistribution to servers or lists, or reuse of any copyrighted component of this work in other works.}
}
\IEEEauthorblockA{\textit{Apple}
}
}

\maketitle

\begin{abstract}
The emergence of large language models (LLMs) has transformed spoken dialog systems, yet the optimal architecture for real-time on-device voice agents remains an open question. 
While end-to-end approaches promise theoretical advantages, cascaded systems (CSs) continue to outperform them in language understanding tasks, despite being constrained by sequential processing latency. 
In this work, we introduce ChipChat, a novel low-latency CS that overcomes traditional bottlenecks through architectural innovations and streaming optimizations. 
Our system integrates streaming (a) conversational speech recognition with mixture-of-experts, (b) state-action augmented LLM, (c)~text-to-speech synthesis, (d)~neural vocoder, and (e) speaker modeling. 
Implemented using MLX, ChipChat achieves sub-second response latency on a Mac Studio without dedicated GPUs, while preserving user privacy through complete on-device processing. 
Our work shows that strategically redesigned CSs can overcome their historical latency limitations, offering a promising path forward for practical voice-based AI agents.
\end{abstract}


\section{Introduction}
The landscape of spoken dialogue systems has undergone a profound transformation with the advent of large language models (LLMs). 
Traditionally, these systems relied on cascaded architectures that combine isolated specialized components---automatic speech recognition (ASR), language understanding, planning, response generation, text-to-speech synthesis (TTS), etc.---into sequential processing pipelines~\cite{glass1999challenges}.
Recently, LLMs are used as the ``interface'' between these cascaded components explicitly~\cite{huang2024audiogpt, an2024funaudiollm}. 
However, researchers and industry practitioners are now pivoting toward a fundamentally different approach: integrating LLMs directly with speech processing capabilities to create comprehensive end-to-end (E2E) speech LLMs that can power conversational agents (CAs) through unified architectures~\cite{nguyen2025spirit,zhang2023speechgpt,tang2023salmonn,defossez2024moshi,fang2025llama,xie2024mini,xu2025qwen2,ji2024wavchat,hurst2024gpt}.

While this shift toward E2E speech LLMs represents a promising direction, it's important to understand the fundamental limitations that drove researchers away from cascaded architectures. 
The sequential nature of processing in cascaded architectures creates barriers to implementing low-latency systems, as each component must wait for complete outputs from preceding stages, thus degrading user experience~\cite{picovoice2024,arora2025espnet}. Additionally, these systems lose important information (e.g., speech characteristics are discarded when transcribing user speech into text) and suffer from error accumulation, where mistakes in earlier components propagate and amplify through the pipeline~\cite{arora2025espnet, xu2025qwen2,chen2024voicebench}. These fundamental limitations explain why the field has embraced E2E approaches that integrate speech and text processing into unified LLM architectures, promising improved quality, lower latency, architectural simplicity, and reduced maintenance complexity.

However, recent research challenges the presumed superiority of and belief in E2E systems. 
Several studies~\cite{nguyen2025spirit,sakshi2024mmau,chen2024voicebench,xu2025qwen2} demonstrate that cascaded systems (CSs) combining state-of-the-art components---such as advanced ASR models with powerful LLMs, or LLMs with a high-quality TTS---consistently outperform their E2E speech LLM counterparts in spoken language understanding and dialog tasks in terms of accuracy.
Furthermore, the quality gap appears to widen in complex scenarios involving multi-turn dialogues and domain-specific knowledge, precisely the areas most critical for practical applications~\cite{arora2025talking}. 
Lastly, CSs are more interpretable as we can track what errors each component makes.

\begin{table}[!t]\centering
    \vspace{-0.1cm}
    \caption{Latency of ChipChat components and subsystems tested with Apple Silicon (Mac Studio, M2 Ultra, 192GB, 2023). Time: $t$ (ms).
    }\label{tab:latency}
    \vspace{-0.3cm}
    \resizebox{0.492\textwidth}{!}{
    \begin{tabular}{cccc}
    \toprule
    & Wait $t$ for input & Inference $t$ per output & Total $t$ up to the point \\
    \midrule
    \faMicrophone{} & 0 & 0 & 0\\
    Mel & 10  & 1 & 11 \\
    ASR & 160 & 13 & 165-175 \\
    LLM State & [pause] & $\sim$ 560 & $\sim560$ \\
    LLM & 0 & 16 & $\sim576$\\
    TTS & [5 words] & 20 & $\sim880$\\
    Vocoder & 25 & 13 & $\sim920$\\
    \faVolumeUp{} & 25 & 0.2 & $\sim920$\\
    \bottomrule
    \end{tabular}
    }
    \vspace{-0.6cm}
\end{table}

The streaming capabilities of E2E systems, once considered their defining advantage, have also come under scrutiny. While theoretically low latency can be provided, many E2E models still exhibit chunking behaviors and processing delays that impact user experience due to non-streaming speech encoders (e.g., Whisper)~\cite{xie2024mini,xu2025qwen2} or speech tokenizations~\cite{mousavi2025discrete}.
This raises fundamental questions: Is it possible to design CSs that (a) maintain their quality advantages while achieving the responsiveness that users expect from conversational agents, and (b) run fully on-device to preserve user privacy?

\definecolor{darkgreen}{RGB}{0, 100, 0}  

\begin{figure*}[th!]
\centering
\begin{tikzpicture}[
    node distance=1cm,
    box/.style={rectangle, draw, minimum width=1.5cm, minimum height=0.9cm, align=center, rounded corners, font=\small},
    io/.style={trapezium, trapezium left angle=70, trapezium right angle=110, minimum width=1.5cm, draw, align=center, font=\small},
    arrow/.style={-{Stealth[length=2.5mm]}, thick}
]

\node[io] (mic) {\faMicrophone{}};
\node[box, right=0.95cm of mic] (mel) {Mel Filterbanks};
\node[box, right=0.95cm of mel] (asr) {Speaker \& \\ ASR};
\node[box, right=1.5cm of asr] (llm) {LLM};
\node[box, right=1.5cm of llm] (tts) {TTS};
\node[box, right=0.95cm of tts] (vocoder) {Vocoder};
\node[io, right=0.95cm of vocoder] (player) {\faVolumeUp{}};

\node[label, below=0.35cm of mic] {\scriptsize \color{darkgreen} Waiting for:};
\node[label, below=0.4cm of mel] {\scriptsize \color{darkgreen} 1 chunk (10ms)};
\node[label, below=0.4cm of asr] {\scriptsize \color{darkgreen} 4-16 frames};
\node[label, below=0.4cm of llm] {\scriptsize \color{darkgreen} 100-400ms pause};
\node[label, below=0.4cm of tts] {\scriptsize \color{darkgreen} 5 words};
\node[label, below=0.4cm of vocoder] {\scriptsize \color{darkgreen} 1 frame};
\node[label, below=0.4cm of player] {\scriptsize \color{darkgreen} 25ms};

\draw[arrow] (mic) -- node[above, font=\scriptsize] {10ms} (mel);
\draw[arrow] (mel) -- node[above, font=\scriptsize] {1 frame} (asr);
\draw[arrow] (asr) -- node[above, font=\scriptsize] {1 token} (llm);
\draw[arrow] (llm) -- node[above, font=\scriptsize] {1-4 tokens} (tts);
\draw[arrow] (tts) -- node[above, font=\scriptsize] {1 frame} (vocoder);
\draw[arrow] (vocoder) -- node[above, font=\scriptsize] {25ms} (player);

\draw[arrow, dashed, color=red] (asr) to[out=30, in=160] (llm);
\draw[arrow, dashed, color=red] (asr) to[out=30, in=160] (tts);
\draw[arrow, dashed, color=red] (asr) to[out=30, in=160] (vocoder);
\draw[arrow, dashed, color=red] (asr) to[out=30, in=160] (player);

\draw[arrow, dashed, color=blue] (player) to[out=200, in=-20] (llm);

\begin{pgfonlayer}{background}
    \node[fit=(mic)(mel)(asr), draw, dashed, rounded corners, fill=blue!5, inner sep=0.2cm] {};
    \node[fit=(llm), draw, dashed, rounded corners, fill=yellow!5, inner sep=0.2cm] {};
    \node[fit=(tts)(vocoder)(player), draw, dashed, rounded corners, fill=red!5, inner sep=0.2cm] {};
\end{pgfonlayer}
\end{tikzpicture}
\vspace{-0.7cm}
\caption{ChipChat's streaming cascaded architecture. Red dashed lines show {\color{red} interruption signals} sent by the ASR to downstream components. The blue dashed line represents the {\color{blue}feedback mechanism} when the audio player sends the feedback information back to the LLM to clear unvocalized generated responses. Text in green shows the {\color{darkgreen} input chunk} that a process awaits before starting its generation.}
\label{fig:chipchat-architecture}
\vspace{-0.5cm}
\end{figure*}
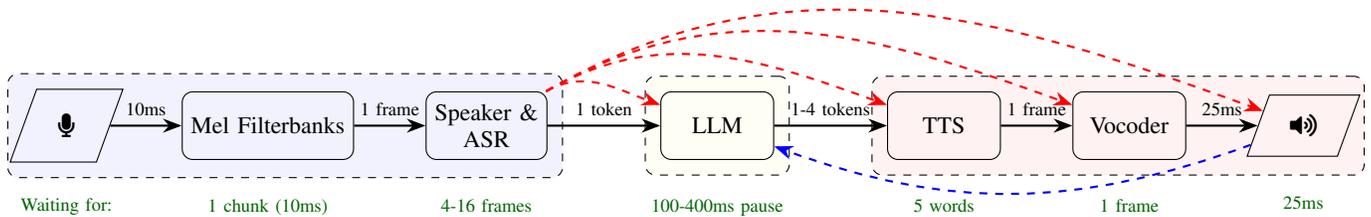

In this paper, we introduce ChipChat, a novel low-latency cascaded system for CAs that directly addresses the discussed historical limitations.
ChipChat is based on: (a) classic design of streaming ASR with mixture-of-experts (MoE) integration, (b) a state-action augmented LLM, (c) a low-latency streaming TTS, (d) a low-latency streaming neural vocoder, and (e) a real-time speaker modeling.
Our approach preserves the performance benefits of specialized components while featuring a streaming design that enables progressive processing and generation across the entire pipeline.
Leveraging MLX~\cite{mlx2023} for components' inference, we are able to achieve sub-second response latency for the entire system on Mac Studio, see Table~\ref{tab:latency}, making advanced CAs more accessible to users while preserving user privacy.
These findings indicate that the future of CAs may not require abandoning cascaded approaches but rather rethinking them to address their inherent limitations.

\section{ChipChat Design}
ChipChat is a distributed system implemented in Python and designed to maximize both memory efficiency and inference speed while minimizing latency. 
We utilize RabbitMQ~\cite{rabbitmq} for inter-process message passing, allowing each component to operate independently while maintaining a cohesive pipeline. 
This design allows each component's output to be streamed immediately upon generation, thus reducing latency.
The system leverages specialized models for ASR, TTS generation, speaker representation, and dialog LLM augmented with state-action. 
An overview of ChipChat is given in Figure~\ref{fig:chipchat-architecture}.

\subsection{Main Components}

{\textit{Microphone Input:}}~~We implement browser-based microphone streaming with near-zero latency audio capture. 
We use the built-in noise cancellation (or voice isolation) capabilities from either the browser or the device hardware and process audio in 10ms chunks to be aligned with the feature computation process while reducing the number of sent messages.

{\textit{Mel Filterbanks}}~~
are computed with 10ms hop distance and 25ms window size and are normalized (zero mean and unit std). Due to normalization we perform running mean and std computation to preserve the streaming property.

{\textit{ASR Model}}~~employs a transformer encoder architecture enhanced with MoE in fully connected layers from~\cite{gu2025asrmoe}. 
The model is trained with Connectionist Temporal Classification~\cite{graves2012connectionist} loss and causal masking to enable streaming inference. 
The model uses 4-frame stacking, 8k word-piece tokenization and 4 experts resulting in $\sim 650$M total parameters. 
To optimize inference speed, we (a) use batched inference (16 input frames found to be optimal); (b) implement key-value caching; (c) introduce a cache resetting mechanism as the training sequence length is less than 30s.
The ASR component also functions as a voice activity detector (VAD) to identify user pauses and turn switches. 
Non-blank, non-repetitive tokens are immediately streamed to the LLM process, where they are encoded right away.

{\textit{Speaker Model}}~~is based on the transformer from~\cite{aldeneh2025speaker} and provides speaker diarization.
We re-train a supervised variant of the model with the same feature extraction as ASR and integrate it directly in the ASR process to leverage its VAD.
At the beginning of the conversation we wait for 3s audio to run the speaker model for the first time and enroll the extracted embedding vector. 
Subsequently we use a 1.5s sliding window, reducing the latency to 1.5s, and call the speaker model only if the current segment contains more than 20\% of speech to avoid model inference for silence segments.

{\textit{LLM:}}~~We employ a dialog LLM enhanced with state-action from~\cite{zhang2025sage}: first, the model infers the user's motivation and emotion, then it predicts the agent's motivation and emotion, and finally it predicts the agent's response.
The model follows an 8x7B Mixtral model (8 experts, 45B total parameters)~\cite{jiang2024mixtral}.
To optimize inference, (a) we pre-encode the conversation prompt in advance before the conversation begins; (b) ASR streams tokens as soon as they are generated, thus the LLM can encode them right away; (c) during LLM turns we pre-encode prompts for the user/agent states in a batch regime without generation; (d) we use the MLX-LM~\cite{mlxlm} key-value rotating cache.
After the LLM generates user and agent states, every generated response token is streamed to the TTS process.

{\textit{TTS model}}~~is based on the streaming architecture SpeakStream from~\cite{bai2025speakstream}.
It uses (a) the streaming speech tokenization via discretization of Mel filterbanks (dMel)~\cite{bai2024dmel}; (b) specially designed interleaving between written $n$-gram words and spoken $n$-gram words.
Such interleaving design enables low-latency generation without loss in the quality.
The TTS employs a transformer decoder (300M parameters) that produces frames at 40Hz. 
To optimize inference, (a) we pre-encode audio prefix used for every sentence generation to better preserve consistency and style; (b) LLM streams tokens as soon as they are generated, thus TTS encodes them right away as they arrive; (c) we implement key-value caching and reset the cache as a new sentence starts.
The system waits for 5 words from the LLM and starts generation afterwards with sending generated tokens immediately to the vocoder.

{\textit{Vocoder:}}~~We use VocStream vocoder from~\cite{bai2025speakstream}.
It is a ParallelWaveGAN model with a fully causal convolutional architecture (13M parameters).
VocStream consists of an upsampler (upsamples 40Hz input to 160Hz) and high-resolution vocoder (converts 160Hz input to the 24kHz wave).
For each incoming frame, the vocoder generates the corresponding audio chunks that are immediately passed to the audio player.

{\textit{Audio player}}~~introduces negligible latency, playing each audio chunk as soon as it becomes available from the vocoder.

\begin{figure}[t]
    \centering
    \includegraphics[width=0.8\linewidth]{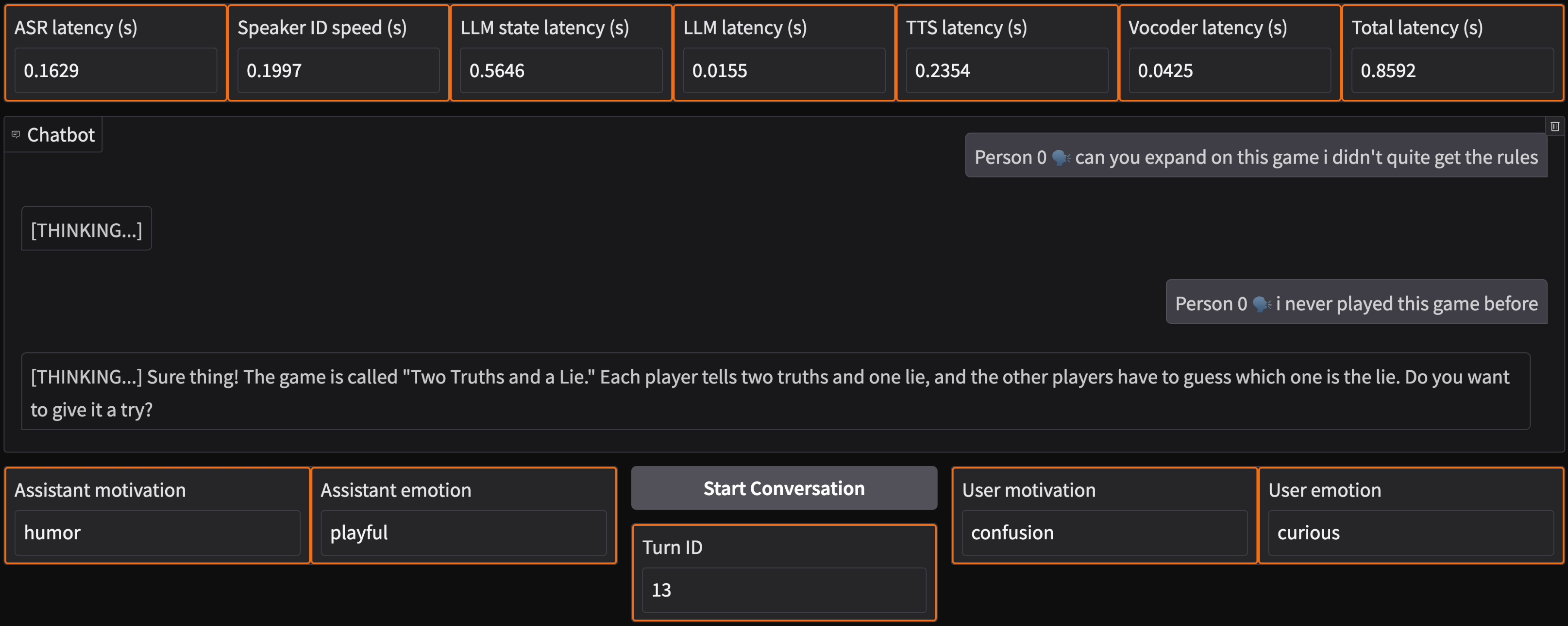}
    \vspace{-0.2cm}
    \caption{ChipChat's Gradio viewer interface.}
    \label{fig:gradio}
    \vspace{-0.6cm}
\end{figure}

{\textit{Viewer:}}~~
We also implement a process with Gradio~\cite{abid2019gradio} to showcase user and agent turns in the text form, user and agent motivation and emotion states predicted by LLM, etc. and to report performance of all components in real time (Figure~\ref{fig:gradio}).

\subsection{Interruption Handling}

Interruption management is a critical aspect of natural conversational systems that has been approached differently.
Some systems, like Moshi \cite{defossez2024moshi}, continue generation despite receiving user input, making interruptions difficult.
Others terminate immediately, potentially creating abrupt transitions~\cite{hurst2024gpt}.

ChipChat implements a responsive interruption system that balances naturalness with responsiveness. 
Our architecture enables immediate detection of user speech through the ASR component, which continuously monitors the audio input. 
When a non-silence token is detected, the ASR process instantaneously signals all downstream components to halt generation (see Figure~\ref{fig:chipchat-architecture}), creating a responsive system that yields the conversational floor when the user begins speaking.

This interruption capability is enhanced by our interruption feedback mechanism: when generation terminates due to user interruption, the audio player sends a signal to the LLM indicating precisely which $n$-gram (leveraging SpeakStream's word-level alignment) was being spoken at the interruption point. 
This allows the LLM to selectively clear its key-value cache of content that was generated but not yet vocalized, maintaining conversational coherence across turns (see Figure~\ref{fig:chipchat-architecture}).
The TTS and vocoder components do not require cache clearing as they reinitialize the cache for each turn. Following interruption, all system components transition to a new conversation turn state, ready to process the user's input.

The interruption sensitivity can be tuned via a configurable parameter that controls the pause duration on which we invoke interruptions. 
Due to continuous listening by the ASR, interruption and its feedback mechanisms, we can set a low value for this parameter to reduce the latency of the system.
Thus our system always monitors whether the user is talking or not and prioritizes user speech over agent speech: if an interruption is detected after the LLM has generated a response that has not yet been vocalized while the user continues to speak, then (a) the system discards the unvocalized generated response; (b) control will be returned to the user turn; (c) the system knows in which state the LLM should be given the feedback mechanism in interruption handling.

\subsection{Context Management and Chunking}

ChipChat implements context management strategies to handle the inherent limitations of fixed-context models while maintaining conversational coherence. 
For the ASR model, we use a temporal chunking that fully resets the model's key-value cache after a configurable period of silence. This approach balances the benefits of contextual information for ongoing speech with the need to avoid limited generalization of positional embedding to longer contexts.
The LLM employs rotating key-value caching. 
By default, the system maintains context for $N$ conversation turns (configurable) before performing a full cache reset.
For the TTS model, we implement sentence-level chunking within a streaming framework. The system monitors incoming text from the LLM for sentence-final punctuation (.!?) and, upon detection, completes the current sentence generation before clearing the key-value cache. This approach ensures prosodic coherence within sentences while allowing for natural breaks between them. To minimize latency during these transitions, we employ carefully designed short prefixes that end with appropriate pauses, ensuring that the first generated frame contains phonetic content rather than silence. The vocoder, being a convolutional model with limited temporal dependencies, does not require explicit chunking strategies, allowing for continuous audio generation across sentence boundaries when appropriate.

\subsection{System Flexibility}
As a cascaded architecture, ChipChat offers flexibility by supporting multiple configurations: ASR-only, ASR and LLM, LLM-only, TTS and vocoder, LLM with TTS and vocoder, vocoder-only, or end-to-end. This modularity facilitates both research experimentation and practical deployment scenarios with varying resource constraints.

\section{Related work and Conclusion}
ESPnet-SDS~\cite{arora2025espnet} is an open-source unified web interface for various cascaded and E2E spoken dialog systems, but runs on a GPU server, and is not designed/optimized for streaming CS.
Recently, Kyutai announced their streaming CS~\cite{kyutai_unmute} although technical details have not yet been published.
Our initial naive implementation of the CS with Pytorch (except the LLM which is still in MLX) and without the aforementioned optimizations had latency more than 4s.

In the paper, we introduce ChipChat, a low-latency cascaded system for conversational agents that overcomes traditional latency limitations while preserving the quality advantages of specialized components. 
ChipChat shows that redesigned parts of cascaded systems with a focus on overall system performance offer a compelling alternative for next-generation conversational agents.
ChipChat achieves sub-second response latency on Mac Studio (Table~\ref{tab:latency}), running entirely on-device. 

\section{Acknowledgments}
We would like to thank: Angelos Katharopoulos, Awni Hanun, Jagrit Digani, Ronan Collobert for help with MLX; Samy Bengio, Christopher Webb, Jamie Cheng for~the~feedback.

\bibliographystyle{IEEEtran}

\end{document}